# Soft x-rays induce femtosecond solid-to-solid phase transition

F. Tavella (1), H. Höppner (2,3,4), V. Tkachenko (5), N. Medvedev (5,6,7) , F. Capotondi (8), T. Golz (2), Y. Kai (3,4) , M. Manfredda (8), E. Pedersoli (8), M.J. Prandolini (9), N. Stojanovic (2), T. Tanikawa (2), U. Teubner (3,4), S. Toleikis (2), and B. Ziaja (5,10)

1) *SLAC National Accelerator Laboratory, Menlo Park, California 94025, USA*

2) *Deutsches Elektronen-Synchrotron DESY, 22607 Hamburg, Germany*

3) *Institut für Laser und Optik, Hochschule Emden/Leer-University of Applied Sciences, 26723 Emden, Germany.*

4) *Institut für Physik, Carl von Ossietzky Universität, 26111 Oldenburg, Germany*

5) *Center for Free-Electron Laser Science, DESY, 22607 Hamburg, Germany*

6) *Institute of Physics, Czech Academy of Sciences, Na Slovance 2, 182 21 Prague 8, Czech Republic*

7) *Institute of Plasma Physics, Czech Academy of Sciences, Za Slovankou 3, 182 00 Prague 8, Czech Republic*

8) *Elettra-Sincrotrone Trieste, 34149 Basovizza, Trieste, Italy*

9) *Helmoltz-Institut Jena, 07743 Jena, Germany*

10) *Institute of Nuclear Physics, Polish Academy of Sciences, Kraków 31-342, Poland*

**Correspondence and requests for materials should be addressed to the following authors:** F. Tavella (e-mail: tavella@slac.stanford.edu), S. Toleikis (e-mail: sven.toleikis@desy.de), V. Tkachenko (e-mail: victor.tkachenko@desy.de), or N. Medvedev (e-mail: nikita.medvedev@fzu.cz)

## Abstract

Soft x-rays were applied to induce graphitization of diamond through a non-thermal solid-to-solid phase transition. This process was observed within poly-crystalline diamond with a time-resolved experiment using ultrashort soft x-ray pulses of duration 52.5 fs and cross correlated by an optical pulse of duration 32.8 fs. This scheme enabled for the first time the measurement of a phase transition on a timescale of ~150 fs. Excellent agreement between experiment and theoretical predictions was found, using a dedicated code that followed the non-equilibrium evolution of the irradiated diamond including all transient electronic and structural changes. These observations confirm that soft x-rays can induce a non-thermal ultrafast solid-to-solid phase transition on a hundred femtosecond timescale.

## Introduction

Femtosecond irradiation of matter with a laser pulse can induce structural transition to a disordered state: amorphization, or defect creation [1–3]. Graphitization of diamond is a counterexample as it is an order-to-order (solid-to-solid) phase transition [4]. Energy delivered to the system by incoming photons leads to a change in the interatomic bonding from $sp^3$ (diamond) to $sp^2$ bonds (graphite). Previous studies showed that graphitization can be achieved with pulses in the regime from optical to hard x-ray energies. Decisive is the average radiation energy absorbed per atom. A dose of ~0.7 eV/atom is needed to graphitize diamond non-thermally [4]. This corresponds to 19.5 GJ/m³ or 5.5 MJ/kg ; a similar damage threshold in terms of the absorbed dose was calculated for optical irradiation of diamond in Ref. [5]. Due to the energy dependence on the photo-absorption cross-sections, achieving such dose per atom requires that the pulse fluence is adjusted to the incoming photon energy.

There are still active debates over the nature of the observed transitions. Ref. [3] claims that for a class of materials, femtosecond optical pulses can trigger a specific process known as a non-thermal phase transition. Generally, for a non-thermal transition, theory predicts that the excitation of a few percent of the valence band electrons leads to a drastic modification of the potential energy surface triggering a displacement of the atoms. This occurs on a much faster time scale (sub-picosecond) than the transfer of the absorbed laser energy to the lattice via electron-phonon coupling (typically few ps) [6,7]. Therefore, such ultrafast transitions are called 'non-thermal', in contrast to the one triggered by atomic heating via electron-phonon coupling which is then referred to as 'thermal'. Contradictory claims can be found in, e.g., Ref. [8].

In addition to these fundamental aspects, understanding the graphitization process is important for diamond-based technologies, since diamond is increasingly used for practical applications in different forms: bulk, thin films [9] or nano-objects [10]. The question of diamond stability and, correspondingly, graphitization, upon high pressure [11,12], annealing [13,14], or irradiation by optical lasers [15,16] have been investigated both experimentally and theoretically.

Recent advent of free-electron lasers (FELs) in the VUV/soft x-ray and hard x-ray regime allows the investigation of a phase transition of diamond on an ultra-short femtosecond timescale. In case of soft x-rays at low pulse fluences the excitation of electrons is only due to single photo-absorption. Photo-absorption of x-rays promotes electrons to highly excited states, which then relax through collisional scattering, producing secondary electrons. This leads to fast electron thermalization during the femtosecond FEL pulse, or shortly after [17]. Because the x-ray cross-section for free carrier absorption is considerably small, there is no direct heating of the free electrons by the x-ray pulses, in contrast to optical pulse irradiation [3]. Therefore x-ray electron excitation, which is driving non-thermal processes, enables an accurate control of the absorbed dose in the material.

Graphitization process of a chemical vapor deposited diamond has been reported in [4] after the irradiation with a VUV/soft x-ray FEL pulse and graphitization was confirmed by post mortem measurements of the irradiated sample area. Photo-absorption cross-section was used to convert the threshold fluence to an average energy absorbed per atom. In work reported in Ref. [4], an average absorbed dose of about 0.7 eV per atom triggers the graphitization process. This threshold value of absorbed energy per atom has been found to be independent of the incoming FEL pulse photon energy in the range between 24 eV and 285 eV. Good agreement of the predicted graphitization threshold with the dedicated theoretical model developed in Refs. [4,18,19] indicated that the observed transition may indeed be non-thermal. However, this experiment did not give any direct evidence on the transition timescale.

Here we report on the experiment in which for the first time soft x-ray induced graphitization of diamond is followed in a time-resolved manner. The experiment was performed at the FERMI free-electron laser facility [20]. A soft x-ray pulse of sufficiently high fluence initiated a structural transition in diamond. The graphitization process was probed by an ultrashort optical pulse. The results from the time-resolved measurement of optical transmittance were compared with the predictions from the theoretical model. They were found to be in a very good agreement, identifying all transition stages anticipated by the model. In particular, the rapid drop of transmission, which is the main signature of graphitization, is clearly visible. It occurs on the predicted timescale of ~150 fs. This time-resolved experimental verification confirms the occurrence of a direct solid-to-solid phase transition induced by a non-thermal modification of the potential energy surface. For different FEL pulse fluences above threshold, the characteristic transmission drop occurs on the same timescale indicating that the same structural transition occurs in all cases. These observations unambiguously confirm that soft x-rays can induce non-thermal direct solid-to-solid phase transition on femtosecond timescale.

**Time-resolved measurement of ultrafast graphitization in diamond**

The method used to observe ultrafast graphitization of diamond is based on solid-state target EUV/optical cross-correlation [22-25]. With this approach, a 47.4 eV FEL pulse is used to induce the transition from diamond to graphite. The experimental setup shown in Fig. 1 is similar to the one used in Ref. [26]. The wavefront of the FEL is tilted with respect to the target; thus the integrated FEL fluence is encoded spatially and temporally onto the surface of the target. The subsequent temporal evolution of the graphitization process is monitored by a probe laser with a wavefront parallel to the target. The experiment was carried out at the DiProI beamline [27] at the FERMI free-electron laser. Details on the experimental setup and beam parameters are available in the Supplementary Materials.

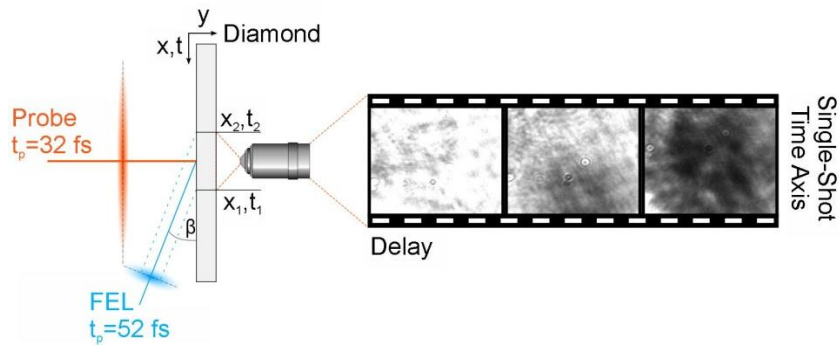

**Figure 1.** Schematic of the experimental setup. The FEL beam propagates at an angle β with respect to the sample surface. Within a single shot, different fractions of the XUV wavefront are absorbed at different spatial positions on the sample, providing a spatio-temporal encoding (Single-Shot Time Axis). A collimated optical pulse under normal incidence is used to probe the optical transmission change. The spatio-temporal encoded signal has a limited time window, which requires additional scanning of the delay between FEL pulses and optical probe pulse.

**Experimental results**

In this experiment, we used a 300 μm thick poly-crystalline CVD diamond substrate which was polished to a surface roughness of less than 20 nm. We performed two specific types of measurements. The first measurement was performed below the graphitization threshold to: (1) measure the FEL pulse duration, (2) define the time overlap between the FEL pulse and the optical probe pulse on the diamond target and (3) measure the threshold for damage (graphitization). The second measurement was performed above the graphitization threshold to investigate the underlying dynamics.

*Below the graphitization threshold*: The FEL focal spot was set to ~6.2x128.1 μm$^2$ (radius at $1/e^2$) to work with lower fluence and a large single-shot time window. The longer spatial dimension corresponds to the time axis. The transient optical cross-correlation provides information on the relative delay between FEL and optical probe pulse. Similar setups are used at FEL facilities to measure the time arrival jitter between FEL and optical probe pulse [23-25].

The characteristic transmission curve of the EUV/optical cross-correlation measured on diamond is displayed in Fig. 2 (blue line).

The measured transient transmission curve is compared to XTANT simulation (solid black line). The time arrival position is marked by the Gaussian pulse (magenta line). The transmission measurement and the simulation are in good agreement, except for the fast relaxation of the signal on a sub-picosecond time scale which we attribute to carrier diffusion (dashed black line). The discrepancy arises as due to the necessary condition of quasi-neutrality within a system with periodic boundary conditions. As simulated in XTANT, only an approximate diffusion model could be applied.

A simpler code from Reference [26] is used to model the pulse duration and the fast signal relaxation. The FEL pulse duration retrieved from single-shot measurements is (52.5±3.4) fs (measurements not shown). These measurements were performed on a $Si_3N_4$ membrane due to its pristine surface quality, which improves the signal-to-noise ratio, using methods described in Ref. [26]. The measurement on diamond shown in Fig. 2a is averaged over 400 measurements. The diffusion of free electron density $n_e$ results in a decrease of carrier density as a function time. The lifetime is estimated to be ~185 fs assuming an exponential decay parameter for $n_e$.

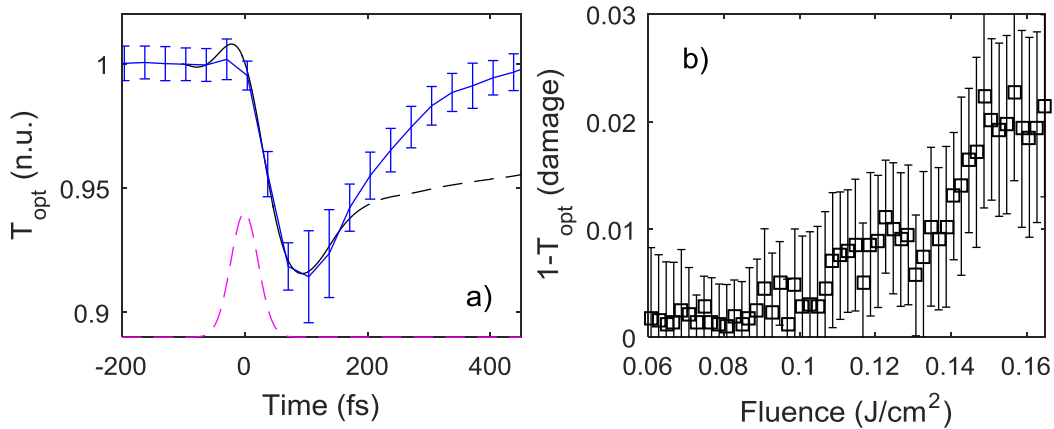

**Figure 2.** Below damage threshold ionization of diamond: a) Transient optical cross-correlation measurement (blue line with error bars) compared to XTANT simulation (black line, solid and dashed, see text). Position of FEL pulse (dashed magenta line). b) Damage threshold measurement: transmission change of optical probe laser versus fluence after 2500 consecutive FEL shots on a diamond sample.

The measurement shown in Fig. 2a was performed below the damage threshold of diamond. The FEL pulse energy on target was (2.29±0.15) μJ. The FEL pulse energy was varied using a gas attenuator. We observed onset of damage on diamond at fluence values >0.1 J/cm² (see Fig. 2b). The peak fluence for this measurement is ~0.17 J/cm², which leads to visible damage after sample irradiation with ~100 shots. The measured damage (graphitization) fluence agree with values reported in Ref. [28] which estimates the damage threshold for CVD diamond at 0.14 J/cm² at 38.1 eV and Ref. [29] with the threshold for graphitization at ~0.1 J/cm² at 50 eV.

*Above graphitization threshold*: Graphitization was performed with a smaller focal spot of ~12.3x17.3 μm² (radius at $1/e^2$) to induce visible graphitization process and to access higher fluences. The smaller focal spot size decreased the single-shot time window. The effect was therefore characterized using single shot data combined with time scans to increase the time window. The FEL pulse energy was (41.67±1.57) μJ, with peak fluence in the focal spot of ~12-18 J/cm² (see Supplementary Materials). We had access to the entire range of fluences due to the spatially encoded measurement (see Supplementary Materials). Data analysis was performed using a sorting method described in the Supplementary Materials.

Each measurement was performed on an undamaged position of the diamond sample. The transmitted probe pulse signal was recorded at a particular time delay with respect to the irradiation with an FEL pulse. In addition, the transmitted probe pulse signal of the post-mortem shot from the previous measurement is recorded on the same CCD image (see Supplementary Materials). Post-mortem analysis of the irradiated sample shows a presence of a graphitized layer on the surface of the irradiated poly-crystalline diamond substrate. From measurements presented in the Supplementary Materials, we conclude that the graphitized layer in the central irradiated region is dominated by nano-crystalline graphite (nc-C) with additional low $sp^3$ amorphous carbon (a-C).

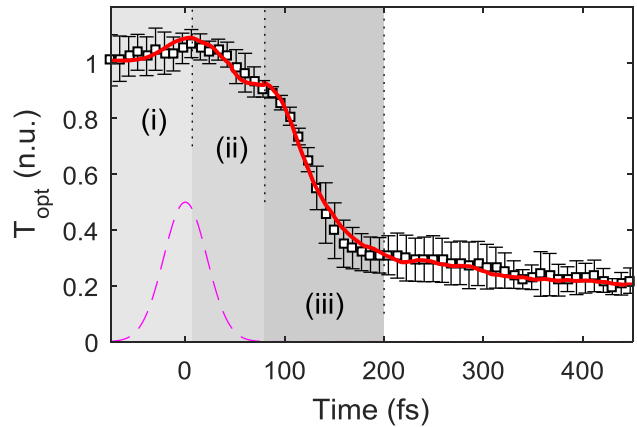

**Figure 3.** Transmission measurement of ultrafast graphitization: The transient optical transmission signal at a wavelength of 630 nm is shown from a diamond sample irradiated with soft x-rays: experimental (open black squares with errorbars) and theoretical with an average dose of 0.8 eV/atom (red line). FEL photon energy was 47.4 eV, pulse duration was 52.5 fs (FWHM, magenta line). Intervals (i)-(iii) denote different stages of graphitization. The predicted thickness of the transient graphitized layer at the time instance of 400 fs is 38 nm.

Comparison between the experimentally measured transmittance of the optical probe pulse (wavelength of 630 nm) with theoretical predictions is shown in Fig. 3. The transmission curve was normalized to the initial transmission of non-irradiated diamond. It exhibits a few characteristic features which reflect three different stages of graphitization: initial electronic excitation, band gap collapse and atomic relocation described in more detail in the following theory section.

**Simulation of ultrafast graphitization of diamond**

We model free-electron laser irradiation of diamond with the in-house simulation tool, XTANT [6,18,19]. It follows the evolution of irradiated diamond through all non-equilibrium stages starting from its x-ray irradiation up to the completion of the graphitization process. Transient information on electronic and atomic structure is obtained from the model. Photo-absorption and non-equilibrium electron kinetics are modeled within a Monte Carlo scheme; low-energy electrons and their coupling to atoms (phonons) are included by means of the Boltzmann collision integral; atomic structure is traced by a molecular dynamics technique, whereas the electronic band structure and the collective atomic potential energy surface for atoms are obtained at each time-step from a transferable tight binding Hamiltonian [21].

Since diamond and graphite have different optical properties, the solid-to-solid phase transition can be monitored by tracing optical parameters such as reflectivity or transmittance. Knowledge of the transient electron distribution, band structure and oscillator strength coefficients (both obtained from the evolving Hamiltonian), fully defines the transient complex dielectric function within the random-phase approximation, from which transient optical coefficients of the materials are derived without any adjustable parameter [30].

In the model, electronic distribution forms the attractive part of the interatomic potential. The bonding between atoms starts to undergo modifications when a sufficient number of electrons is excited. This can lead to an ultrafast non-thermal phase transition. In case of diamond, our model predicts that if an average dose of above 0.7 eV per atom in diamond is absorbed, it leads to the excitation of over 1.5% of electrons from the bonding states of the valence to the antibonding states of the conduction band. Accounting for a wide band gap in diamond (~6 eV), the excitation of 1.5% of electrons produces sufficiently large changes of the potential energy [31]. This, in turn, triggers non-thermal graphitization [4,19]. For more details, see Supplementary Materials.

XTANT predicts that non-thermal graphitization of diamond proceeds in a few steps: first, the electrons emitted after FEL photoabsorption excite a sufficient number of electrons into the conduction band via collisional processes. For photon energies around 50 eV, it can already occur during a femtosecond pulse [17]. Second, the high number of electrons excited to antibonding states weakens interatomic bonds, which leads to a band gap collapse [4]. This occurs within ~50 fs after the pulse maximum. Third, the band gap collapse promotes even more electrons into the conduction band, weakening the interatomic potential within the diamond lattice. Interatomic bonds are then breaking and forming $sp^2$ bonds instead of $sp^3$. Finally, diamond graphitization completes within the next 50-100 fs. Due to the small mass of carbon atoms and the short distance needed for their relocation to the new equilibrium positions within the forming highly-overdense graphite planes, the entire non-thermal phase transition completes within ~150 fs after the pulse peak.

**Simulation results: stages of graphitization**

Non-thermal graphitization of diamond proceeds in following steps (see Fig. 3):

(i) Initial electronic excitation occurs during the FEL pulse. In case of VUV or soft x-ray pulses, photoelectrons relax to the bottom of the conduction band within a few femtoseconds via collisional processes and the resulting electron cascade [17,18]. A density of the electrons excited to the conduction band is shown in Fig. 4. The simulated increase of the transmission during the FEL pulse is due to the initial electronic excitation, altering the optical coefficients. For FEL photons of 47.4 eV energy with small penetration depth (~0.02 μm) and for relatively low density of excited electrons, our model predicts a rapid (small) decrease of the reflectivity, corresponding to an increase of transmission (cf. both coefficients plotted in Fig. 7 in Supplementary Materials). The experimental uncertainty in this regime does not allow for accurate comparison of the data with theoretical results.

(ii) Electronic excitation triggers a band gap collapse (see Fig. 5). This occurs within ~50 fs after the pulse maximum of the FEL pulse at the time instant when the density of conduction band electrons overcomes the threshold value of ~1.5% [19,31]. The band gap collapse is reflected in the slight decrease of the transmission, followed by a short-lived plateau (up to ~ 80 fs). The short-lived plateau corresponds to a delay between the band gap collapse and the start of the sufficient atomic relocation to the new equilibrium positions within overdense graphite planes [19].

(iii) The significant (steep) decrease of the transmission occurs at times 80-150 fs after the FEL pulse maximum. It is followed by the atomic relocation (occurring at ~140-200 fs), which significantly changes the material properties, from insulating diamond (Fig. 6a) to semi-metallic graphite (Fig. 6d). The electronic density in the conduction band further increases (the second rise in Fig. 4), leading to the final atomic relocation [4]. In this rearrangement, atoms settle at the new positions corresponding to overdense graphite with broken plane orientations (Figs. 6c-d) [19].

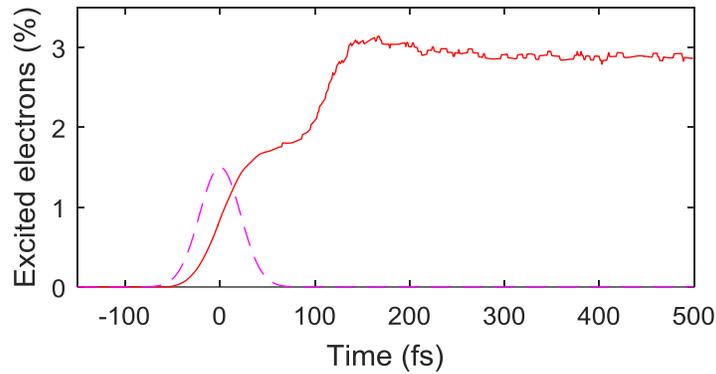

**Figure 4.** Calculated percentage of electrons excited to the conduction band (red line) after irradiation of diamond with a FEL pulse of 47.4 eV photon energy and 52 fs pulse duration (FWHM, magenta line). Average absorbed dose was 0.8 eV/atom.

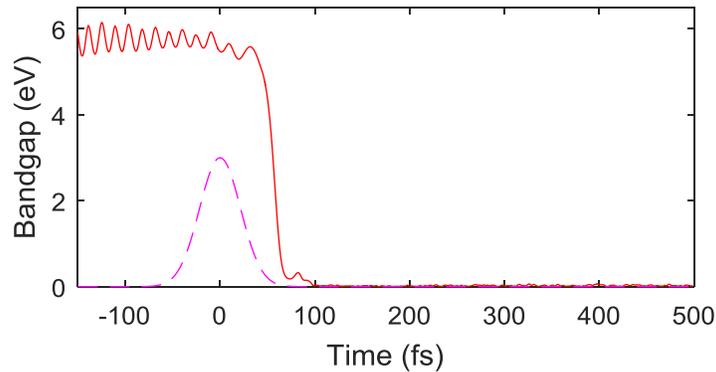

**Figure 5.** Calculated band gap (red line) after FEL irradiation of diamond with a FEL pulse of 47.4 eV photon energy and 52 fs pulse duration (FWHM, magenta line). Average absorbed dose was 0.8eV/atom.

The simulation results clearly indicate that the ultrafast graphitization occurs within ~150 fs after the exposure to a femtosecond soft x-ray pulse. Corresponding changes in the optical transmittance are shown in Fig. 2 for below- and in Fig. 3 for above-threshold absorbed doses. In the below-threshold case (0.6 eV/atom), which does not induce graphitization, there is only a slight visible drop of the transmission curve, quickly recovering to its original value due to electron particle and carrier diffusion (Fig. 2). For any under-damage threshold dose, the transmission curve never drops further below ~0.9. In case of the above-damage threshold dose, there is always a second, major drop. Thus, the occurrence of such a drop can be unambiguously interpreted as an experimental signature of the phase transition. After reaching the overdense graphite state at ~200 fs, the optical transmission stabilizes at the value of ~0.3 (Fig. 3) for another few hundreds fs.

We would like to emphasize that the induced phase transition is of non-thermal nature. It is caused solely by the modification of the interatomic potential and not by atomic heating via electron-phonon coupling. Our model predicts that the electron-phonon coupling acts at picosecond scales (as also known for other materials [31]). During the graphitization process reported here (which occurs within ~150 fs), the contribution of heating by electron-phonon coupling is only minor. Moreover, we have verified that if in our theoretical model we 'switch off' the electron-phonon coupling, the results remain the same: ultrafast graphitization without electron-phonon coupling occurs within the same time span of ~150 fs. This again confirms the non-thermal transition scenario [31,4].

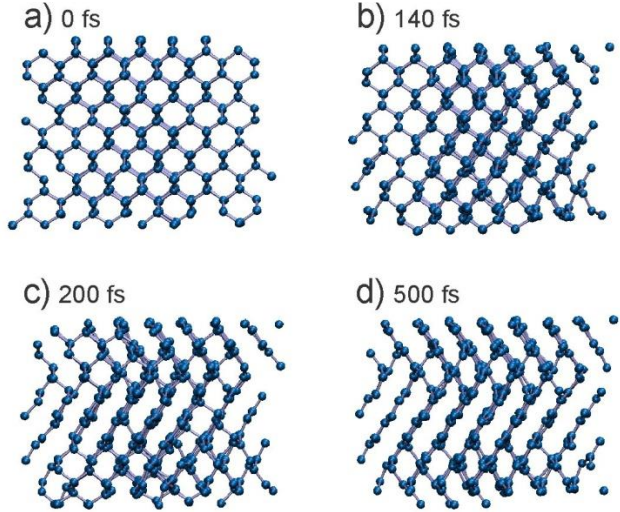

**Figure 6.** Calculated atomic snapshots at different time instances after irradiation with soft x-ray pulse of 47.4 eV photon energy, 52 fs duration (FWHM) at the average absorbed dose of 0.8 eV/atom. The number of atoms in the simulation box is 512. The graphitization is delayed with respect to the FEL pulse, with t=0 fs the FEL pulse maximum. Plot (a) shows diamond at t=0 fs. Structural changes start to occur at (b) t=140 fs and continue until (c) t=200 fs, i.e., the rearrangement of atoms from diamond to graphite structure takes ~ 60 fs. Afterwards, the graphite formed remains almost unchanged until the simulation stops at (d) t=500 fs.

For different FEL pulse fluences, the corresponding transmission drop occurs on the same timescale (see also Supplementary Materials) indicating that the same phase transition occurs in all cases. The predicted stages (especially (ii) and (iii)) are visible in the experimental curves (see Fig. 3 and Supplementary Materials). They are in a good agreement with theoretical predictions within the measurement accuracy given by statistically estimated error bars. The timescales of the predicted processes match the experimental results with high accuracy. In particular, the rapid drop of transmission, which is the main signature of graphitization, is clearly visible and occurs on the predicted timescale (between 100 fs and 150 fs from the FEL pulse maximum).

This timescale is much shorter than any known timescale of electron-phonon coupling and thermal phase transition, which are usually expected to occur on a picosecond to nanosecond timescale [7,21,32]. Moreover, it is even faster than any non-thermal melting (solid-to-liquid) transitions which require at least some 300-500 fs, for example, in silicon, gallium arsenide, and other covalently bonded semiconductors [1,3,33]. The non-thermal melting in bismuth was claimed to occur within 190 fs in Ref. [34]. However, this timescale was estimated from the exponential decay time of a diffraction peak. Here we define the timescale as the time between the latest time instant when transmission was still equal to 1, and the time instant at which the transmission's rapid decrease stops (slope change at ~ 200 fs). Thus, applying the same criterion to Bi, we would obtain the corresponding melting timescale of at least 500 fs (c.f. Fig. 3c in [34]). Therefore, to the best of our knowledge the non-thermal phase transition described in the present work is much shorter than any previously reported.

## Summary

Non-thermal graphitization of diamond occurring on a 150 fs timescale has been observed experimentally. The time-resolved pump-probe experiment was performed at the FERMI FEL facility. The transient transmission of a 300 µm thick diamond substrate pumped with 47.4 eV, 52 fs (FWHM) soft x-ray pulses from the FERMI FEL was probed by 630 nm, 32.8 fs (FWHM) optical laser pulses. High temporal resolution of the experimental scheme made it possible to follow all stages of the soft x-ray induced graphitization. The results from the time-resolved measurement of the optical transmittance were compared with the predictions from a theoretical model. They were found to be in excellent agreement within the measurement accuracy. The characteristic transmission drop occurred on the same timescale for different FEL pulse fluences above threshold, indicating that the same structural transition took place in all cases. This time-resolved experiment confirms for the first time the occurrence of a direct solid-to-solid ultrafast phase transition induced by a non-thermal modification of the potential energy surface. This finding is important both for fundamental research and for technological applications of diamond as a nanomaterial.

## Acknowledgements:

The authors thank the technical staff of FERMI facility at Trieste for their support during the experiment. The authors are especially grateful for the support of the laser group at FERMI. The authors thank J. Gaudin, H. O. Jeschke, Z. Li, R. Santra and O. Vendrell for helpful discussions.

## Author contributions:

F.T., S.T., N.M., B.Z. conceived the experiment.
F.T., S.T., M.P., Y.K., U.T. M.M. E.P, H.H., N.S., T.G, T.T. and F.C were responsible for the experimental set-up and performed the experiment.
F.T., H.H. analyzed the data.
V.T., N.M., B.Z. performed theoretical simulations.
F.T., S.T., V.T., N.M., B.Z. wrote the manuscript with discussions and contributions from all other authors.

## Additional information:

Authors declare no competing financial interests.

## Supplementary Materials for

## "Soft x-rays induce femtosecond solid-to-solid phase transition"

**Details on theoretical model**

Time evolution of optical coefficients for x-ray irradiated diamond was modeled with the hybrid code XTANT [1,2]. The code uses tight-binding molecular dynamics to follow positions of atoms within a supercell, and to calculate transient electronic structure within the valence band and within the bottom of the conduction band. These transient data on the atomic and electronic structure establish potential energy surface $\Phi(\{\mathbf{R}(t)\}, t)$ that changes with time,

$$\Phi(\{\mathbf{R}(t)\}, t) = \sum_{j=1}^{N} f_e(E_j) E_j(t) + E_{rep}(\{\mathbf{R}(t)\}), \quad (1)$$

where the first term is an attractive part formed by the electrons. Electrons are described with the electron distribution function $f_e(E_j)$ (electron occupation numbers normalized to two, accounting for the spin degeneracy). They populate the transient energy levels $E_j$, which are the eigenstates of the transient Hamiltonian $H(\{\mathbf{R}(t)\})$ obtained by its diagonalization. The electronic contributions are summed up over all energy levels N. The second term describes the repulsive part: $E_{rep}$ is the core-core repulsive potential of ions with $\{\mathbf{R}(t)\}$ which is a set of atomic coordinates for all atoms in the simulation box.

The Newtonian equations of motion, with the force derived from the potential energy surface, Eq. (1), are solved for all atoms in the simulation box with periodic boundary conditions using the Verlet algorithm in its velocity form. A typical calculation uses 512 atoms in the simulation box.

As one can see from Eq.(1), the construction of the potential energy surface requires knowledge of the transient electron distribution, $f_e(E_j)$. In order to trace the electron distribution function, XTANT splits electrons into two energy fractions: (i) high-energy electrons of the energy above a certain threshold (~10 eV; the choice of this threshold value was analyzed in Ref. [1]) which are modeled as individual particles with Monte Carlo (MC) event-by-event simulation scheme [1], and (ii) low-energy electrons of the energy below the threshold, which are assumed to be in local thermodynamic equilibrium and obey Fermi-Dirac distribution. They are modeled with Boltzmann equation [2]. Boltzmann collision integral for the energy exchange with ions (electron-phonon coupling) reads as [3]:

$$I_{e-at}^{TB} = \frac{2\pi}{\hbar} \sum_{f=1}^{N} |M_{e-at}(E_i, E_f)|^2 \begin{cases} f_e(E_i)[2 - f_e(E_f)] - f_e(E_f)[2 - f_e(E_i)]g_{at}(E_i - E_f), \text{for } i > f \\ f_e(E_i)[2 - f_e(E_f)]g_{at}(E_i - E_f) - f_e(E_f)[2 - f_e(E_i)], \text{for } i < f \end{cases}, (2)$$

where $g_{at}(E)$ is the integrated atomic distribution (the integral of the Maxwellian distribution with a transient ion temperature), and $M_{e-at}(E_i, E_f)$ is the matrix element describing the scattering of an electron on the atomic displacement during the current time-step [3].

When an electron loses its energy below the low-energy threshold, it joins the electronic low-energy fraction, thereby heating up the low-energy electronic distribution. Photoabsorption and decays of K-shell holes (if produced), are also incorporated within the MC scheme. High-energy electrons perform secondary ionizations, exciting new electrons from the valence to the conduction band of diamond. Probabilities of such impact ionizations are calculated from the scattering cross sections, which are derived from the transient complex dielectric function [3,4]. The details of the MC scheme including the scattering cross sections were reported in Ref. [1].

Low-energy electrons form a Fermi-Dirac distribution on the transient energy levels, i.e., band structure, obtained by the diagonalization of the transient tight binding Hamiltonian mentioned above. The Hamiltonian depends on the position of all atoms in the simulation box via overlap integrals (in the Slater-Koster form [5]). These so-called hopping integrals are functions of interatomic distances and positions, properly adjusted to reproduce atomic structure of various carbon-based materials. Thus, the method is capable of reproducing different phase transitions and equilibrium structure of both, diamond and graphite, and also a number of other phases such as linear chain of carbon atoms (down to carbon dimer), SC, BCC, HCP and FCC structures although with poorer agreement [6]. The parametrization of hopping integrals, the construction of the tight-binding Hamiltonian, the expressions for the forces, and a numerical scheme used in the tight-binding molecular dynamics calculations are described in detail in Ref. [1].

For analysis of the diffusion effects we extended the XTANT code to account for the heat transport in the electronic subsystem by means of a rate equation, following the work [7].

For the analysis of diffusion effects we use an approximate model applicable under the restriction of periodic boundary conditions. Accordingly, we extend the XTANT code to account for the energy transport in the electronic subsystem by means of a rate equation, following work in Ref. [7]:

$$\frac{\partial T_{e,a}}{\partial t} = \frac{T_0 - T_{e,a}}{\tau}, \qquad (3)$$

where $T_0$ is the temperature of the unexcited surrounding bath (room temperature), and $\tau$ is the characteristic relaxation time, chosen equal to be 500 fs. This time corresponds to the characteristic evolution time of the affected layer thickness discussed below. The heat diffusion effects only play a role for an under-threshold energy deposition, whereas for the above-threshold doses the structural material damage is so fast that it is practically unaffected by the diffusion-induced energy sinks. The only effect of diffusion is then shifting of the damage threshold dose to higher values, by roughly a factor of 50%, since now a part of the energy is being lost.

In the presented application of our model for soft x-ray induced graphitization of diamond, the incoming soft x-ray pulse is assumed to have a Gaussian temporal profile, with pulse parameters corresponding to the experimental ones. This particular choice of the fixed temporal pulse shape for the description of single shot data from (SASE) FEL experiment is justified as in Ref. [8]. Here we showed that for the pulses of the same duration graphitization depends on the pulse fluence, and not on its specific temporal profile. A Gaussian temporal pulse distribution is also a good assumption for the seeded FERMI FEL, since it reflects the seeding laser temporal distribution. The photon energy was 47.4 eV, corresponding to the experiment. The pulse duration was 52 fs (FWHM). Pulse fluence varied around the damage threshold estimated to be ~0.7 eV/atom, ranging from a sub-threshold (0.6 eV/atom) to the above-threshold values (~1 eV/atom).

Using electron diffusion equation, with the diffusion coefficient taken from Ref. [9], we verified that already a fluence of ~5 J/cm$^2$ quickly produces almost homogeneous electron density in a few-tens-nanometer thick layer of diamond corresponding to a near-threshold dose. A fluence of >20 J/cm$^2$, gives a maximum absorbed dose of much over 1 eV/atom near the surface, which then relaxes to lower values due to the fast electron transport. These estimations allow us to use a fixed absorbed energy per atom for the evaluation of the transmission signal with our model while varying only the thickness of the excited layer.

Complex dielectric function (CDF), dependent on the optical wave frequency $\omega$, is calculated within the random-phase approximation (RPA) [10]:

$$\varepsilon^{\alpha\beta}(\omega) = \delta_{\alpha\beta} + \frac{e^2 \hbar^2}{m_e e_0 \Omega} \sum_{nn'} \frac{F_{nn'}}{E_{nn'}^2} \frac{f_e(E_n) - f_e(E_{n'})}{\hbar\omega - E_{nn'} + i\gamma} \qquad (4)$$

where $\Omega$ is the volume of the simulation box; $E_{nn'} = E_{n'} - E_n$ is a transition energy between two eigenstates $E_n$ and $E_{n'}$ within the band structure; $f_e(E_n)$ and $f_e(E_{n'})$ are the corresponding transient electron distribution functions; $m_e$ is the mass of a free-electron; $e$ is the electron charge; $\hbar$ is the Planck constant; and $e_0$ is the vacuum permittivity in SI units. Broadening parameter is set to $\gamma = 1.5 \times 10^{13}$ s$^{-1}$, however, the particular choice of $\gamma$ does not affect the results beyond the broadening of peaks in the CDF [10]. $F_{nn'}$ is the oscillator strength corresponding to the energy transition and is defined as $F_{nn'} = \langle n|\hat{p}|n'\rangle$ a momentum matrix element between the two eigenstates, which can be obtained from the tight binding Hamiltonian [10].

The calculated CDF gives the transient values of the complex refractive index:

$$\tilde{n}(\omega) = n + ik = \sqrt{\varepsilon(\omega)} \qquad (5)$$

which is then converted into the reflectivity ($R$) and transmission ($T$) coefficients for an incidence angle $\theta$ (which in our experiment is set to 90°) [9]:

$$R = \left| \frac{\cos\theta - \sqrt{\tilde{n}^2 - \sin^2\theta}}{\cos\theta + \sqrt{\tilde{n}^2 - \sin^2\theta}} \right|^2 \qquad (6)$$

$$T = \left| \frac{4\cos\theta \sqrt{\tilde{n}^2 - \sin^2\theta} \exp\left(-i\frac{2\pi d}{\lambda}\sqrt{\tilde{n}^2 - \sin^2\theta}\right)}{\left(\cos\theta + \sqrt{\tilde{n}^2 - \sin^2\theta}\right)^2} \right|^2 \qquad (7)$$

Note that there is no fitting parameter involved in calculation of the complex dielectric function and optical coefficients. The optical transmission coefficient also depends on the optical wavelength, $\lambda$, and transient thickness of the excited layer of diamond during its phase transition to graphite, $d$ [11]. Initially non-irradiated diamond is transparent to the probe pulse of 630 nm used in the experiment. The soft x-ray FEL-photons of 47.4 eV have an attenuation length of 26 nm [12].

The thickness of the affected layer evolves from the initial value of 26 nm to a transient graphite thickness estimated from data at $t$=400 fs. It changes in time according to the heat diffusion and hot-carrier transport. We model these transport effects in an approximate way by assuming an average diffusion. Thus, the thickness of the affected layer, used for calculation of the transmission, is evolving in time with the typical square-root dependence ~ $(t/\tau)^{1/2}$. The results shown in the plots are convolved with the finite-duration probe pulse, which is assumed to have Gaussian temporal profile of 32.8 fs duration (FWHM corresponding to the experiment). This convolution smoothes the results, but does not affect the timescales.

**Measurement details**

A 300 μm thick poly-crystalline CVD diamond substrate was mounted on a motorized XYZ

stage. The sample was illuminated by FEL radiation at an angle of $\beta = 20°$ with respect to its surface. We used an FEL beam with a photon energy of ~47.4 eV (pulse energy (41.67±1.57) µJ, pulse duration (52.5±3.4) fs). The propagation direction of the optical probe laser pulse was normal to the sample surface. The optical probe laser had pulse energy of ~0.3 µJ on target and a pulse duration of 32.8 fs measured with autocorrelation. The center wavelength of 630 nm was chosen as we expected a higher probe contrast in the graphitization effect when compared to longer wavelengths. The transmitted probe laser pulse was imaged with a Mitutoyo™ ×10 microscope objective onto a CCD camera (see Fig.1). A large aperture probe beam of ~1.7 mm diameter at $1/e^2$ was used to homogeneously illuminate the region of interest, similarly to experiments performed in Ref. [13]. This type of illumination provides a constant spatial intensity distribution of the probe beam on the imaged sample region of interest. This prevents the need to calibrate the measurement data with spatially varying probe laser intensity distribution.

Two different types of experiments are performed below and above the graphitization threshold, using the beamline adaptive optical system [14] to change the FEL beam size, and the gas attenuator to vary the FEL pulse energy.

(i)    Below graphitization threshold and FEL calibration measurements

These measurements were performed at low fluence and with a large single-shot time-encoded spatial dimension with a measurement window of several hundreds of femtoseconds. This setup is used to measure the FEL pulse duration, measure the time delay between FEL and optical probe pulse (time arrival measurement) and to assess the threshold for graphitization. An advantage of using a seeded free-electron laser for this experiment is the excellent arrival time stability of the FEL pulse in respect to the optical probe pulse, compared to non-seeded FELs [15]. During this experiment we have also measured the arrival time and the FEL pulse duration, using the same experimental setup on a thin $Si_3N_4$ membrane before the graphitization experiment. The FEL pulse duration is (52.5±3.4) fs and the time arrival jitter is below 5 fs rms for short term measurements (100 shots). Details on this particular setup can be found in References [13].

(ii)   Graphitization measurement

Details on the experimental setup and method used to measure ultrafast graphitization of diamond can be found in Ref. [13]. This method is based on solid-state target EUV/optical cross-correlation. In Ref. [13] it was used for single shot measurements. In our case, the measurement window defined by the spatial dimension of the FEL focus was less than 60 fs and additional scanning of the pump-probe delay was necessary to observe all stages of graphitization process. Fig.1a shows a typical transmission measurement of the 630 nm probe pulse penetrating through a diamond sample irradiated with an FEL pulse. Figs. 1b and c show the transmission of the probe pulse through FEL irradiated diamond recorded at different times and the corresponding post mortem products.

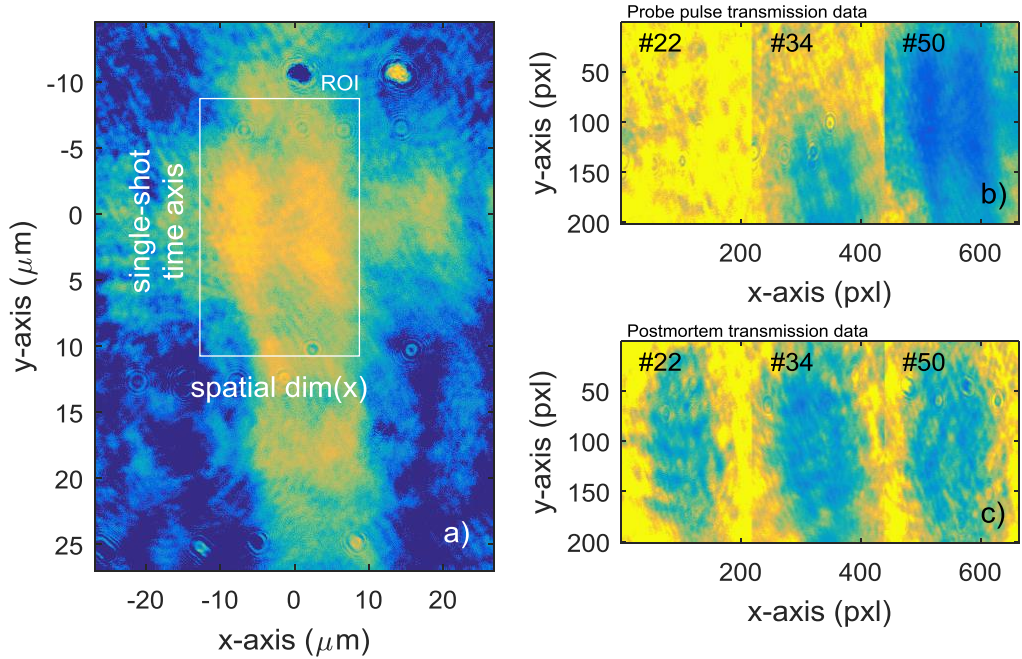

**Figure 1.** Examples of raw data used for the reconstruction of the graphitization process:. a) absorption measurement representing the focal spot of the FEL beam ($\lambda_{FEL}$ = 26.17 nm). Region ROI of ~200 x200 pixel was used for the evaluation of the data; b) example of absorption data shots; and c) corresponding absorption data of the post-mortem shots. Insets depict scenarios when: #22 - probe pulse hits the target prior to the FEL pulse; #34 - probe pulse hits the target approximately at the same time as the FEL pulse; #50 - probe pulse hits the target after the FEL pulse.

Shot #22 was recorded at a time when the probe pulse hits the target prior to the FEL pulse. #34 shows arrival of the probe pulse and FEL pulse at approximately the same time. A fast transmission drop is visible, which corresponds to the onset of the graphitization process in a single shot. Shot #50 was recorded after the main graphitization process was over. The scanning delay between shots is 10 fs, which corresponds to a relative delay between (#22 and #34) and (#34 and #50) of 120 fs and 160 fs, respectively. The single shot information encoded in the time axis (y-axis in Fig.1) is also used to reconstruct the graphitization process. In order to determine the spatial-to-temporal conversion factor, the imaging system was calibrated, using a 228 lines/mm grating placed at the sample plane. The spatio-temporal calibration was calculated using the crossing angle β and was estimated to be 0.305±0.007 fs/pxl. This value was confirmed by two time scans, resulting in (0.3061±0.0032) fs/pxl.

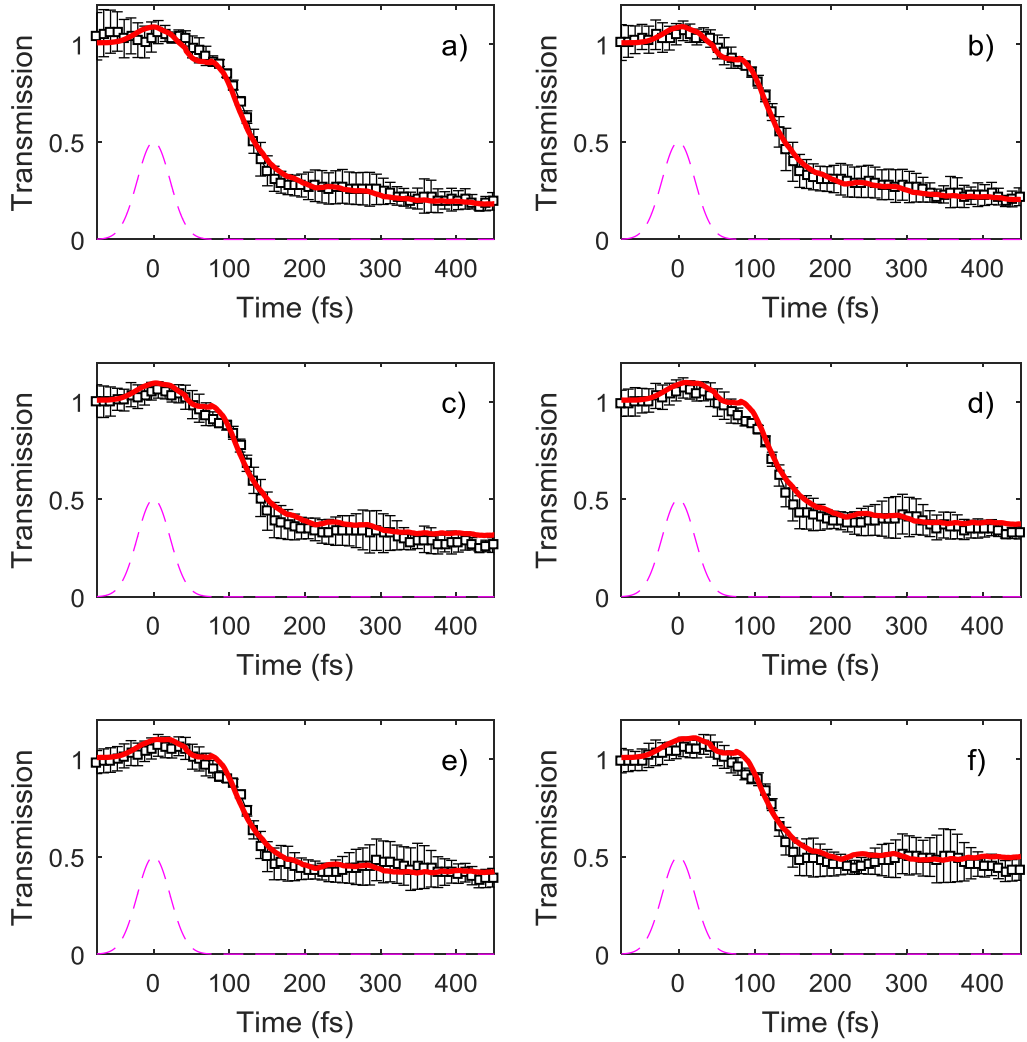

**Figure 2**. Transient optical transmission signal from soft x-ray irradiated diamond: experimental (black squares) and theoretical ones (red curves) for 630 nm probe pulse. FEL photon energy was 47.4 eV. Pulse duration was 52 fs (FWHM of Gaussian profile, magenta line). Data and predictions were obtained for various pulse fluences producing transient graphite layers of the following thickness at the time instance of 400 fs: a) 40 nm; b) 38 nm; c) 29 nm; d) 25 nm; e) 22 nm; f) 17 nm.

We measured the time evolution of the carrier dynamics in diamond, probed with the 630 nm pulses. On each frame we moved the sample to a new spatial position on the diamond substrate in order to acquire the new data shot. On the same frame we acquired the optical transmission of the postmortem shot at the previous measurement position, thus having the postmortem shots available for analysis. We proceeded as follows to sort the measurement data. We used postmortem shots to sort the data from the time scans to find data points in a number of various absorption ranges. Equal or similar amount of absorption in the postmortem shots corresponds to equal amounts of graphite present on the sample. The data was 'binned' between 1 and 0.2 with 0.02 grid steps. "0" corresponds to a total signal transmission (no graphitization). The corresponding data in the measurement shot was sorted, using the information obtained from the post-mortem shots. This corresponded to a direct selection of data with graphitized layer of a fixed width. The data shots taken at different pump-probe delays and sorted using post-mortem probe pulse transmission values were used to generate time-dependent evolution plots of the graphitization process for different pulse fluences. Plots for different soft x-ray pulse fluences were compared with the theoretical predictions in Fig. 2. In each case the transient thickness of the affected layer was assumed to relax to the value equal to the thickness of the graphitized layer at t = 400 fs (as described above). This value was obtained from the measured transmission value, using transient optical coefficients obtained from our simulation at this time instant.

**Post mortem sample analysis**

After the single-shot FEL/optical cross correlation experiments, post mortem analysis of the samples was carried out in order to determine the presence of graphite layers on the surface of the irradiated polycrystalline diamond substrates (300 μm thick, grown by chemical vapor deposition (CVD)). During the single-shot FEL/optical cross correlation experiments, the sample was moved after each FEL pulse, so that the next pulse would radiate a fresh part of the sample. An example optical microscope image of an irradiated diamond sample is shown in Fig. 3a. For future reference, we compare three areas: region A is an area exposed to the center of the FEL pulse; the FEL was not perfectly Gaussian and region B covers an area exposed to a side lobe of the FEL; and region C was an area not irradiated by the FEL (Fig. 3a).

During the post mortem analysis, a number of techniques were used to characterize and confirm the presence of graphitization. The thickness of the graphite layers was measured with an ellipsometer as well as using optical transmission methods. Scanning probe microscopy (SPM), also known as atomic force microscopy (AFM), was used to map the surface morphology of the sample and gain information on the surface roughness and sample consistency. Confocal Raman spectroscopy was used because this method is highly sensitive to the microstructure, demonstrating unique experimental signatures between diamond, nano-crystalline graphite (nc-C) and amorphous carbon (a-C). And finally, x-ray photonelectron spectroscopy was used to measure the electronic configuration (diamond – $sp^3$, graphite – $sp^2$) of the carbon atoms. Using the last two methods, the composition of the graphitized layer in the central irradiated region A is dominated by nc-C with additional low $sp^3$ amorphous carbon (a-C).

*Thickness measurements:* Different graphitized layer thicknesses and surface morphologies resulted from irradiation with different FEL fluences on a particular sample area. In region A (Fig. 3a), the peak FEL fluence was estimated to be ~12-18 $J/cm^2$, well above the ablation threshold of carbonaceous materials. The lower estimate is calculated from transmission measurements (see Fig. 1a) as (12.22±0.46) $J/cm^2$. An upper value estimate of 17.7 $J/cm^2$ is calculated assuming a Gaussian focal distribution. The surrounding area contains several side lobes of graphitization with thinner layers. These lobes are caused by diffraction lobes on the FEL beam and were exposed to lower fluence levels. Because of ablation during FEL radiation, it is not possible to obtain a clear correlation between the graphitized layer during the transient state (<1ps after irradiation) and the measured post-mortem thickness. However, in order to have a complete picture of the process, we have measured the post mortem thickness with optical transmission methods with an optical probe laser (630 nm center wavelength). The graphite thickness is maximum at the central region of A, with a value of <70.1±11.7 nm using the complex refractive index of a-C to estimate an upper limit for the average thickness [16]. In addition, an average thickness (including side lobes) of the graphitized sample was carried out with an ellipsometer (UVISEL phase modulated spectroscopic ellipsometer, Horiba Scientific) and was estimated to be 34.07±3.02 nm. The resolution of this method is limited due to the large laser spot size of 1 mm at $1/e^2$ and is also dependent on the ellipsometric model.

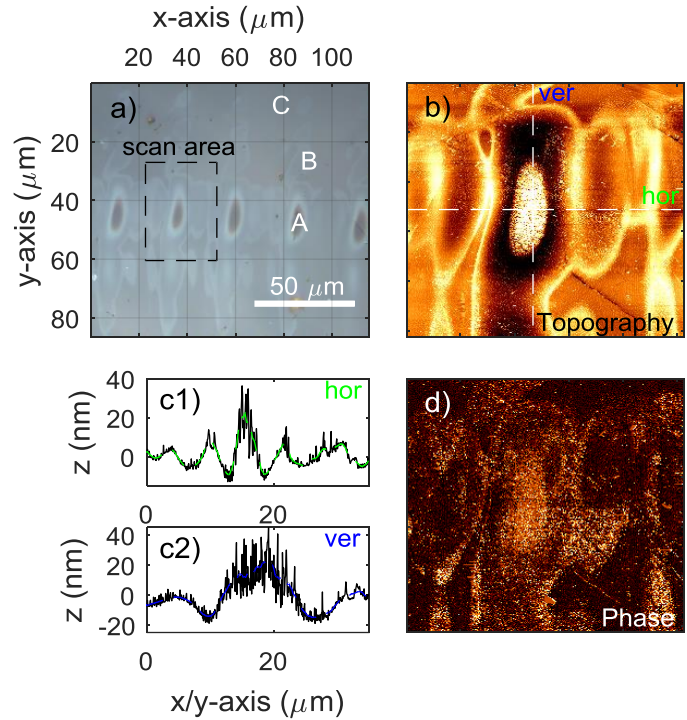

**Figure 3:** SPM sample measurements: a) Optical microscope image of a series of single-shot graphitization damage. Each spot was produced upon irradiation with a single FEL pulse. The image also shows the position and size of the SPM scanned area (dashed box). Regions A, B and C denote measurement areas for confocal Raman measurements (shown in Fig. 4); b) Topography of a 35x35 µm$^2$ non-contact SPM scan. c1) and c2) are vertical and horizontal lineouts from (b). The scan is normalized to the diamond substrate surface (zero line is the diamond substrate surface); d) Phase scan indicates areas with softer material as brighter zones. Darker zones are a harder material, such as the diamond substrate on the top right-hand side of the figure.

*Scanning Probe Microscopy (SPM, Park XE-70)*: SPM was used to generate a high resolution topographical surface map of part of the irradiated sample with area of 35x35 µm$^2$. The chosen SPM scanned area is shown in the optical image of Fig. 3a (x100 magnification microscope image). In the resulting SPM topographical map (Fig. 3b), the central spot is caused by the main pulse of the FEL and some side lobes are also observable as lighter areas. Horizontal lineouts from the SPM map are shown in Fig. 3c(1) and c(2). The central spot and some of the side lobes are clearly observable in the lineouts. From this measurement, evidence of ablation is visible, as well as formation of craters around the central spot and the side lobes. This effect may be attributed to the expansion during graphitization not only in sample normal direction (z), but also in the sample plain. It is worth noting, that we do not observe typical circular features around the central spot which would indicate melt expulsion induced by the recoil pressure of the ablating material (so-called piston-effect). In the phase scan (Fig. 3d), brighter zones are areas with softer material which we associate with carbonaceous material. Darker zones are harder material, such as the diamond substrate on the top side of the figure.

*Confocal Raman Micro-Spectroscopy (XploRA series, Horiba Scientific)*: The confocal Raman measurements were performed in a backscattering geometry using x100 infinity corrected objective, focusing a 532 nm probe laser to a ~1µm$^2$ spot. Raman spectroscopy at this probe wavelength is only sensitive to sp$^2$ bonds, because they have a 50-100 times larger scattering cross section compared to sp$^3$ bonds [17].

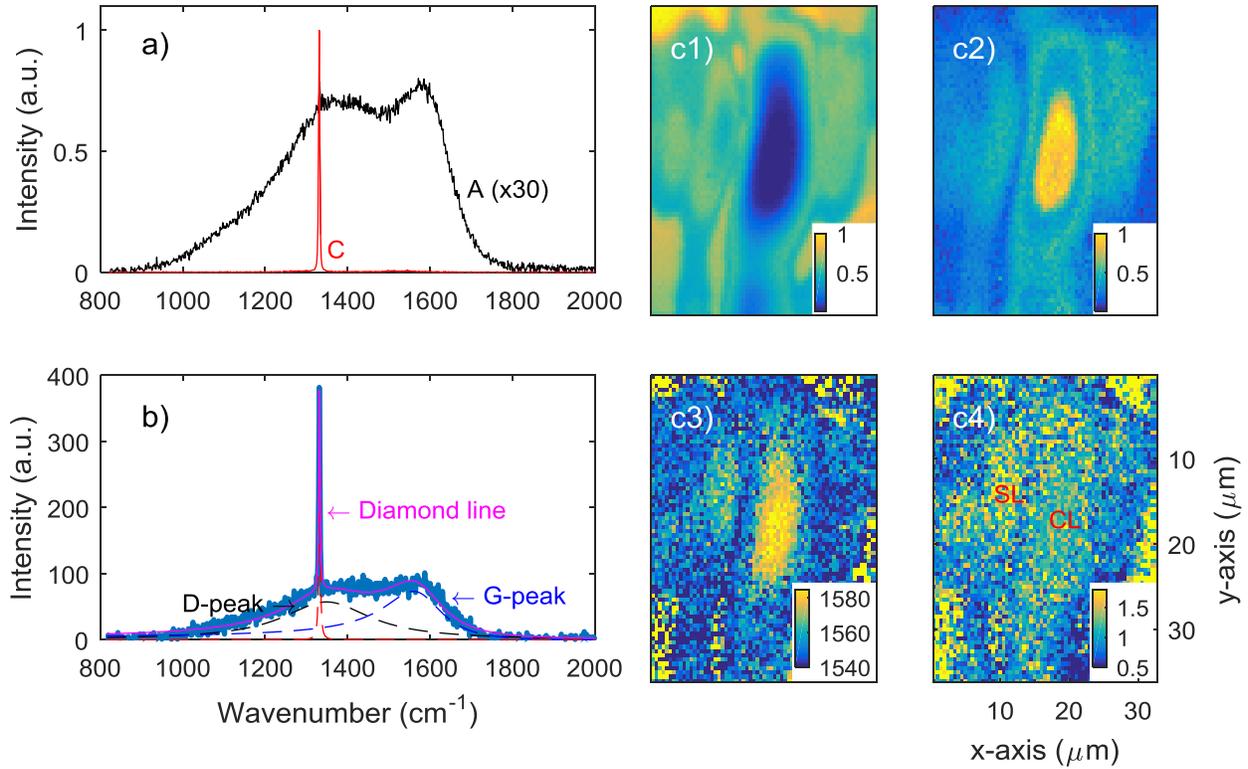

**Figure 4**: Confocal Raman microscopy measurements: a) two spectra are shown; at the center of region A (black) and in region C (red). b) A single spectrum is shown at a side lobe position B (blue). A, B and C denote measurement positions shown in Fig. 3a and all spectra were taken with a spatial resolution of 1μm$^2$. In addition, b) demonstrates the decomposition of a complex line shape: diamond-peak at 1332 cm$^{-1}$ (magenta), D-peak at ~1355 cm$^{-1}$ (dashed black) and G-peak at ~1575 (dashed blue) [18]. Using the spectral decomposed b), the images were analyzed into the following spatial maps: c1) spatial distribution of the diamond component; c2) the spatial distribution of the combined sum of G-peak and D-peak; c3) the spatial map of the shift of wavenumber of the fitted G-peak; c4) the spatial map of the area ratio between D and G peaks.

Confocal Raman spectra at two different positions were taken at center of region A of an irradiated spot, and region C at a non-irradiated position (Fig. 4a). The black line in Fig. 4a has two distinct characteristic peaks: G-peak at ~1575 cm$^{-1}$ and a D-peak at ~1355 cm$^{-1}$. This spectrum is similar to the measured spectrum of graphitized diamond irradiated by a FEL at 24 eV (Fig. 3 in [19]). In addition, this spectrum is typical for nano-crystalline graphite (nc-C) compared to amorphous carbon (a-C) because of the presence of the distinct G- and D-peaks [18,19]. The red line in Fig. 4a is a pure diamond line at 1332 cm$^{-1}$. For comparison, a confocal Raman spectrum was taken at region B (caused by a side lobe of the FEL at lower fluence) and is composed of nc-C/a-C and diamond (Fig. 4b). This figure also demonstrates how a Raman spectrum can be decomposed into a diamond-, D- and G-peak. The diamond component (dotted magenta line) was fitted with a Lorentzian curve, while the graphitized layer spectra can be decomposed in two characteristic peaks: D-peak (black) and G-peak (blue). This method is widely used to characterize nc-C/a-C films [20,21]. The D- and G-peaks are fitted with a Lorentzian and a Breit-Wigner-Fano (BWF) curve, respectively [18].

Using the spectral decomposition shown in Fig. 4b, a confocal Raman microscope scan was carried out on a 32.5x36 μm$^2$ sample area with 1 μm$^2$ resolution. The results of this scan are shown in Fig. 4c. The diamond-, D- and G-peak are fitted to the measured Raman spectrum at each position (scan step 0.5 μm in x/y sample surface direction), to determine composition maps of the irradiated area. Figs. 4c1 and 4c2 represent the concentration maps of diamond and the nc-C/a-C components. There is a complete absence of diamond at the main central spot position A (Fig. 4c1). In contrast in Fig. 4c2, a large nc-C/a-C component can be found at the central spot A on the sample. A wavenumber shift of the Raman G-peak to lower frequencies indicates a higher $sp^3$ fraction [22,23]. Thus observed shift of the wavenumber of the G-peak indicates that nc-C dominates the composition at the central position A, where most FEL energy was deposited [21], compared to position B. For example, the G-peak position at the center A is 1578.2±5.09 cm$^{-1}$ and at B 1562.8±3.45 cm$^{-1}$ at the side lobe location (Fig. 4c3). Confirmation of a dominate nc-C component of the central spot can be determined from the fitted area ratio from the D- and G-peak (D/G-ratio, see Fig. 4c4). Graphite is made of stacked graphene layers. The in-plane size of these layers, and therefore the size of the nano-graphite crystallites, can be denoted by a parameter *La*, and has been shown to be related to the area ratio of (D/G) [24]. Also according to Ref. [25], the $sp^3$ fractions are inversely proportional to the integrated (D/G) ratio. The (D/G) ratio is 1.15±0.21 at the center spot (A) and 1.42±0.39 at the side lobe position (B). Therefore the estimated size of *La* is ~3.8 nm at A and 3.1 nm at B using the analysis given by [24]. The high strain energy at the diamond/graphite interface is also responsible for this size limitation [19].

*X-ray Photoelectron Spectroscopy (XPS, PHI-5000 VersaProbe Scanning XPS System)*: This XPS system uses an Al (K-α) source (1486 eV). The measurement depth is 10-50 Å and the spot size used during the measurement is 9x9 μm$^2$. XPS can be used to measure the binding energy of the carbon atoms and discern the hybrid $sp^3$ and $sp^2$ bonds and the diamond peak [20]. The carbon 1s peak position in diamond is 285.50 eV, about 1 eV higher than the 1s core-level binding energies of the $sp^3$ and $sp^2$ hybrids (sp2 carbon ~284 eV and $sp^3$ carbon ~284.9 eV). The binding energy of the $sp^3$ hybrids is shifted with respect to the sp2 hybridized carbon by ~0.9 eV. The carbon 1s peak can be de-convoluted into two sub-peaks and the ratio of the fitted curves can be used to calculate the $sp^2/sp^3$ ratio on the irradiated sample area [26].

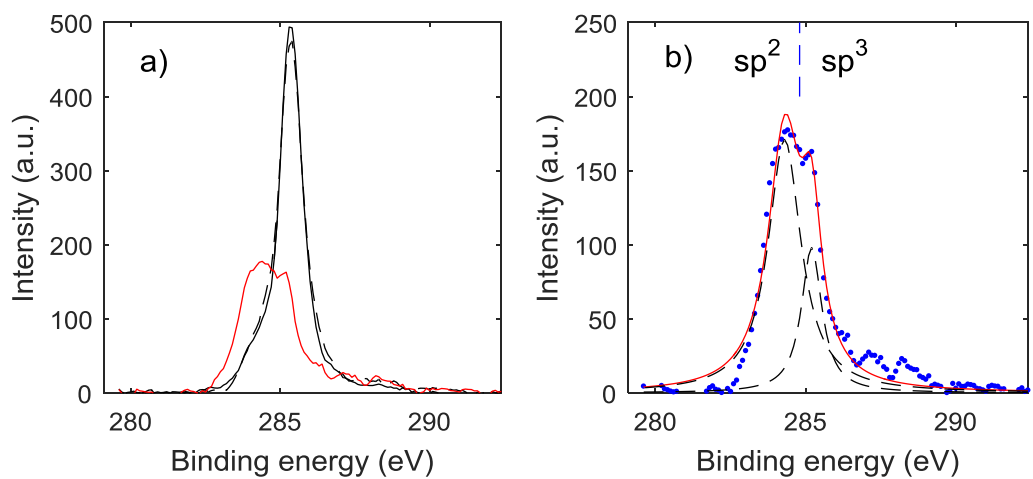

**Figure 5**: Photoemission spectra from XPS measurements and analysis: a) Measurement of a non-irradiated diamond sample at position C (black line) and measurement of an irradiated sample at position A (red line). b) Fitting procedure to determine sp2/sp3 concentration at position A: XPS data (blue dots); dashed black lines are the Lorentzian line fits of the sp2 and sp3 hybrids and the sum of these lines (red line).

The photoemission spectra are taken at two different sample positions A and C (see Fig. 3a). At position C, the non-irradiated sample revealing a diamond peak centered at ~285.5 eV (black line Fig. 5a). Due to the relatively large spot size of the XPS x-ray beam compared to the central spot at A, the graphitized sample area could only be coarsely scanned, shown in (red line in Fig. 5a). This spectrum was fitted with two Lorentzian lines representing the $sp^2$ and $sp^3$ hybrids, as described in Ref. [26] (see Fig. 5b). The analysis results in sp$^3$ content of ~26.6%.

**Simulation results**

Simulations with different transient layer thicknesses (central value of 38±4 nm) are shown in Fig. 6a. They are compared to the experimental results shown in Fig. 2b. This comparison gives an estimate for a 'fit' error during adjusting the transient layer thickness in the simulation to the experimental data.

Further tests indicate that the evolution of the affected layer thickness produces only a minor effect at the considered FEL fluences and at the timescale of non-thermal graphitization. In fact, as shown in Fig. 6b, after varying the characteristic evolution time of the affected layer by one picosecond, the results remain almost unaffected. In particular, the change does not affect the major transmission drop corresponding to the non-thermal phase transition.

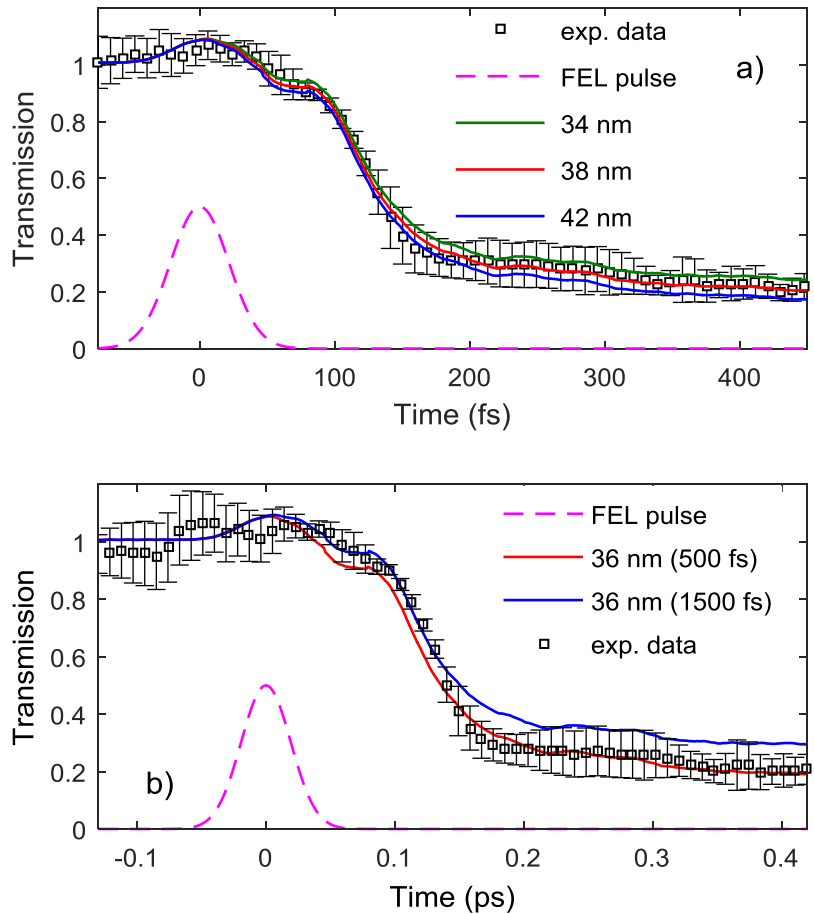

**Figure 6.** Transient optical transmission signal from soft x-ray irradiated diamond: experimental (squares) and theoretical ones (green, red and blue curves) for 630 nm probe pulse. FEL photon energy is 47.4 eV. Pulse duration is 52 fs (FWHM of Gaussian profile, magenta line). Data and predictions are shown for: (a) various pulse fluences producing transient graphite layers of the thickness between 34 nm (green line) and 42 nm (blue line) at t=400 fs. The central line corresponds to the transient layer thickness of 38 nm (red line), and b) evolution of affected layer of 36 nm transient thickness with the time constant of 500 fs (red line), and its evolution with the time-constant of 1500 fs (blue line).

Figure 7 demonstrates transient behavior of all three calculated optical coefficients: transmission, reflection and absorption. For the below-damage threshold dose (Fig. 7a), the transmission and reflectivity drops, and the absorption increased during the FEL pulse due to the excitation of electrons to the conduction bands. These values are preserved until the excited electrons are relaxed at a picosecond timescale. Accounting for the heat diffusion, as described above, by means of Eq. (3), would allow the transmission values to restore closer to its original value. However, the present model is too simplistic to produce a quantitative agreement. Our focus here is on the above-threshold doses and the induced phase transition. The below-threshold irradiation can be investigated in future with dedicated tools. For above-threshold absorbed dose (Fig. 7b), atomic rearrangement changes the band structure, which strongly affects the optical properties, greatly decreasing the transmission and increasing the reflectivity as well as absorption, in a two-step process. For example, the reflectivity initially decreased during the laser pulse due to excitation of electrons to the conduction band but then increases.

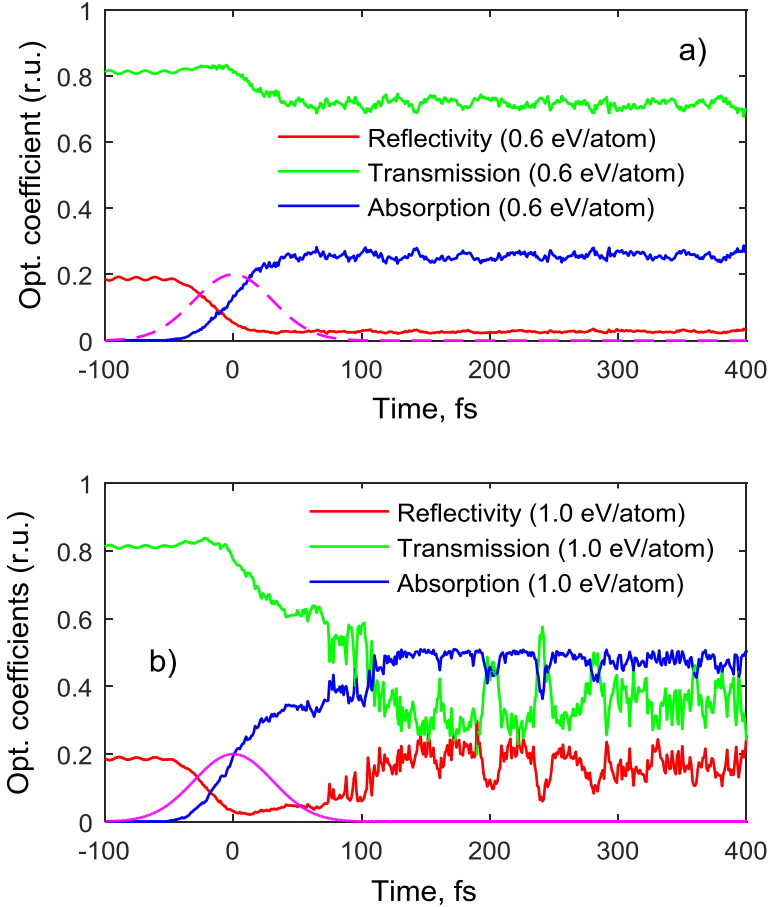

**Figure 7.** Predicted transient optical coefficients: transmission, reflection and absorption in irradiated diamond for below- and above-threshold doses of a) 0.6 eV/atom, and b) 1.0 eV/atom, respectively. FEL photon energy is 47.4 eV; pulse duration is 52 fs (FWHM of Gaussian profile, magenta line). Probe pulse with 630 nm photons was assumed.

(Make a new paragraph)

For the above threshold dose, XTANT also predicts that electrons after FEL irradiation are excited to high-energy states, which, in case of the Fermi FEL laser parameters means energies of a few tens of eVs. In contrast to hard X-ray irradiation (cf. e.g. Ref. [2]), such electrons relax within a few femtoseconds [27] to the low-energy states at the bottom of the conduction band of diamond. Fig. 8 illustrates that for both doses, below and above the damage threshold, the number of excited electrons just follows the FEL laser pulse profile. Secondary excited electrons are of low energy, and are not capable of exciting other electrons.

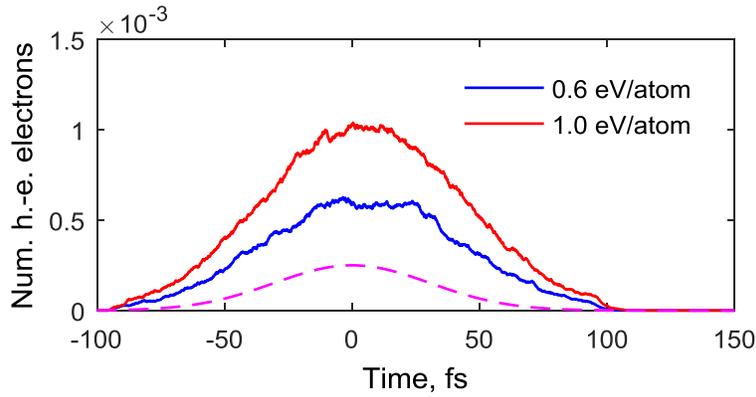

**Figure 8.** Predicted transient density of high-energy electrons in irradiated diamond for below- and above-threshold doses. FEL photon energy is 47.4 eV; pulse duration is 52 fs (FWHM of Gaussian profile, magenta line).

The total number of electrons in the conduction band (both, high- and low-energy fractions) is shown in Fig. 9. The figure confirms that the number of conduction band electrons generally raises during the FEL pulse, but its further evolution differs in below-damage and above-damage cases. In case of below-damage threshold, the number of electrons stays the same, until the electrons later cool down and recombine which in diamond occurs on the scale of ~200 ps [28]. This long-timescale process is not included in our model. For the above-damage threshold case, there is a second increase of the conduction band electron density, taking place at around 100 fs during the band gap collapse, see Fig. 10.

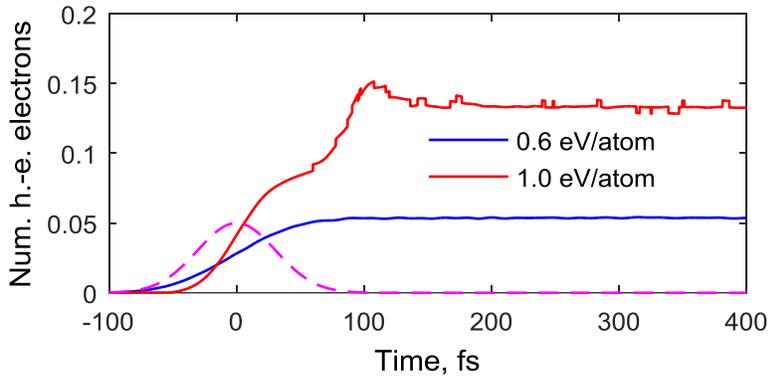

**Figure 9.** Predicted transient density of all electrons excited into the conduction band in irradiated diamond for below- (0.6 eV/atom) and above-threshold (1.0 eV/atom) doses. FEL photon energy is 47.4 eV; pulse duration is 52 fs (FWHM of Gaussian profile, magenta line).

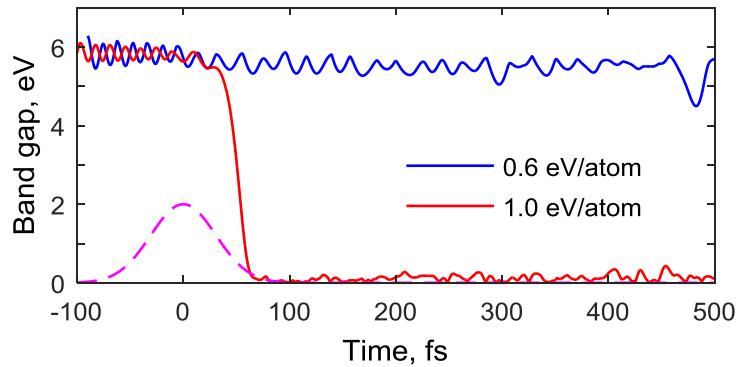

**Figure 10.** Predicted transient band gap of irradiated diamond for below- (0.6 eV/atom) and above-threshold (1.0 eV/atom) doses. FEL photon energy is 47.4 eV; pulse duration is 52 fs (FWHM of Gaussian profile, magenta line).

The band gap collapses during the graphitization [1], when the atoms start to relocate to the new positions within the graphite planes, changing its bonding from $sp^3$ to $sp^2$. This process is 'self-amplifying': as the number of excited electrons reaches ~1.5%, it triggers the band gap collapse. The shrinking of the band gap promotes even more electrons into the conduction band. When their density reaches around 3%, the phase transition becomes irreversible, and all atoms settle down in their new equilibrium positions within the (overdense) graphite planes. The nearest neighbor distance in diamond is 1.53 Å, whereas in graphite it is 1.41 Å. These distances differ only by a fraction of angstrom. The interplane graphite distance at femtosecond timescales is very short, the distance expands only later to bring graphite layers into its equilibrium density.

The initially excited electrons in the conduction band have a very high temperature on the order of 20000 K (see Fig. 11). This is the threshold temperature, corresponding to the excitation of ~1.5 % of electrons across the band gap from the valence into the conduction band. Note that this is close to the temperature of nonthermal melting in silicon (~17000 K), which promotes ~9% of electrons to the conduction band, due to the much smaller band gap in Si (1.17 eV). For above-threshold doses, electron temperature decreases during the phase transition, as atoms are gaining kinetic energy while relocating to the new minimum of the potential energy corresponding to graphite. However, such an increase of the atomic temperature and corresponding decrease of the electronic temperature is *a consequence*, not the cause, of the nonthermal graphitization, as it was discussed in detail in Ref. [1].

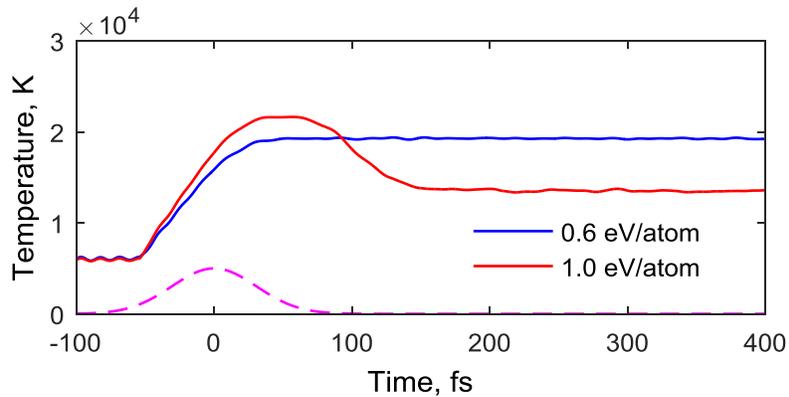

**Figure 11.** Predicted transient temperature of electrons in irradiated diamond for below- (0.6 eV/atom) and above-threshold (1.0 eV/atom) doses. FEL photon energy is 47.4 eV; pulse duration is 52 fs (FWHM of Gaussian profile, magenta line).